\documentclass[aps,pre,amsmath,amssymb,twocolumn]{revtex4}
\usepackage{amsmath,amsthm,amssymb}
\usepackage{graphicx}
\usepackage{pstricks}
\usepackage{enumitem}
\usepackage{xcolor}
\usepackage{amsmath,amsthm,amssymb}
\newcommand{\be}{\begin{equation}}
\newcommand{\ee}{\end{equation}}
\newcommand{\ba}{\begin{eqnarray}}
\newcommand{\ea}{\end{eqnarray}}
\newcommand{\baa}{\begin{eqnarray}}
\newcommand{\eaa}{\end{eqnarray}}
\newcommand{\ed}{\end{document}}
\newcommand{\lab}[1]{\label{#1}}
\newcommand{\re}[1]{(\ref{#1})}
\newcommand{\ci}[1]{\cite{#1}}

\begin{document}

\title {Dirac particles  on periodic quantum graphs}
\author{J.R. Yusupov$^{1}$, K.K. Sabirov$^{2,3}$  and D.U. Matrasulov$^{1,4}$}
\affiliation{$^1$Yeoju Technical Institute in Tashkent, 156 Usman Nasyr Str., 100121, Tashkent, Uzbekistan\\
$^2$Tashkent University of Information Technologies, 108 Amir
Temur Str., 100200, Tashkent Uzbekistan\\
$^3$Tashkent State Technical University named after Islam Karimov, 2 Universitet Str., 100095, Tashkent, Uzbekistan\\
$^4$Turin Polytechnic University in Tashkent, 17
Niyazov Str., 100095, Tashkent, Uzbekistan}

\begin{abstract}
We consider the Dirac equation on periodic networks (quantum graphs). The
self-adjoint quasi periodic boundary conditions are derived. The secular
equation allowing us to find the energy spectrum of the Dirac particles on
periodic quantum graphs is obtained. Band spectra of the periodic quantum
graphs of different topologies are calculated. Universality of the probability
to be in the spectrum for certain graph topologies is observed.
\end{abstract}
\maketitle

\section{Introduction}
Quantum graphs have attracted much attention as an effective tool for modeling
particle and wave dynamics in branched quantum structures. The advantage of
quantum graphs in modeling quantum transport in low-dimensional branched
structures and networks comes from the fact that the description can be
effectively reduced into a one-dimensional Schr\"odinger (Dirac) equation on
metric graphs, which can be exactly solved in most of the cases.  Quantum
graphs are determined as the branched quantum wires, which are connected to
each other at the nodes (vertices). Wave and particle dynamics in such systems
are described in terms of quantum mechanical wave equations on metric graphs.
Initially, quantum mechanical treatment of a particle motion in branched
structures was considered in quantum chemistry of organic molecules. References
\ci{Pauling,Rud,Alex}, where  the electron motion in branched aromatic
molecules was studied, can be considered as a pioneering attempt for the study
of particle motion in quantum graphs. However, the strict formulation of the
quantum graph concept, where the latter was defined as a branched quantum wire,
has been presented a few decades later by Exner and Seba in \ci{Exner1}.
Further progress was made by Kostrykin and Schrader, who proposed general
vertex boundary conditions providing self-adjointness of the Schr\"odinger
equation on quantum graphs \ci{Kost}. Later the quantum graph concept has been
used in different contexts (see,
Refs.\ci{Uzy1,Kuchment04,Uzy2,Gaspard,Exner15,Grisha,Barra,Uzy3,Mugnolo,Uzy4,Bolte1,Hul,Jambul,PTSQGR})
and an experimental realization in microwave networks was done \ci{Hul}.
Despite considerable progress made on the study of different aspects of quantum
graph theory, most of the studies are limited by considering the
nonrelativistic case, i.e., unlike its nonrelativistic counterpart, the study
of relativistic quantum dynamics on graphs is still remaining out of focus in
quantum graphs theory. The first, who treated Dirac equation on graphs, were
Bulla and Trenkler in \cite{Bulla}. Later this problem has been considered in
\ci{Bolte,Harrison}, where strict formulation of the problem with the
self-adjoint vertex boundary conditions were presented and spectral and
scattering properties were studied. However, the research in this topic did not
get further development. In \ci{Jambul2020}, the problem of reflectionless
quasiparticle transport in quantum graphs were studied. In \ci{KarimBdG}, the
Bogoliubov-de Gennes equation on graphs modeling Majorana fermions in branched
quantum wires were considered. Physical systems, which can be modeled in terms
of the Dirac equation on periodic quantum graphs appear e.g., in polymer
physics. The well known Su-Schriefer-Heeger (SSH) model
\cite{SSH,Heeger2,Heeger3} describing polaron dynamics in conducting polymers,
such as, e.g., polyacethylene in its continuum version leads to the Dirac type
equation \cite{Campbell1981,Campbell1982}. When one considers so-called
(periodically) branched conducting polymers, polaron transport in such
structures can be described in terms of the Dirac equation on periodic quantum
graphs \cite{Exciton,Polaron}. Synthesis and study of electrophysical
properties of such  branched polymers have been reported recently in the
literature \cite{BCP14,BCP15,BCP16,Chernyak43}. Another system where the Dirac
equation on periodic quantum graphs can be applied comes from optics, where the
optical emulation of one-dimensional (1D) Dirac fermions is possible
\cite{Alex1,Alex2}.

In this paper, we address the problem of the Dirac equation on periodic quantum
graphs. In particular, we present a model for Dirac quasiparticles in branched
lattices, which can be mapped on to the periodic quantum graphs. We note that
different  aspects of the Schr\"odinger equation on periodic quantum graphs
have been studied in detail in
Refs.\ci{Grisha1,PQG1,PQG2,PQG3,PQG4,PQG5,PQG6,PQG06,Rabinovich,Grisha2,Grisha4,PQG7,Grisha20,PQG8}.
Some mathematical properties of the continuum and discrete Schr\"odinger
operator on graphs are studied in \ci{PQG5,PQG6,PQG06,PQG7}. An effective
numerical method for the determination of the spectra of the Schr\"odinger
operator on periodic metric graphs is presented in \ci{Rabinovich}. Physically
acceptable models of quantum graphs and a very effective approach for the
treatment of their band spectra and dispersion relations have been proposed in
\ci{Grisha}. Quantum transport in periodic quantum graphs is considered in a
very recent paper \ci{PQG8}. It is important to note that during the past
decade, the problem of wave dynamics in networks has been successfully extended
to the case of the nonlinear wave equation (see, e.g.,
Refs.~\ci{SGEEPL,Adami17,dimarecent,Karim2018,Chsol,Exciton,BJJEPL} and
references therein). The tight binding approach for studying band spectra of
periodic quantum graphs is developed in \ci{Grisha20}. Bethe-Sommerfeld
conjecture $\mathbb{Z}^2$-periodic graphs are studied in \ci{PQG2}.

The paper is organized as follows. In the next section we briefly recall the
Dirac equation on quantum graphs. Section III presents the boundary conditions
and general secular equation for the Dirac equation on periodic quantum graphs.
In Sec. IV we compute and study the band spectra of the periodic graphs of
different topologies. In Sec. V the probability that a randomly chosen momentum
belongs to the spectrum of the periodic graph is investigated. Finally, Sec. VI
presents some concluding remarks.

\section{Dirac equation on quantum graphs}
As mentioned above, quantum graphs are determined as one- or quasi-one
dimensional branched quantum wires, where the wave dynamics can be described in
terms of quantum mechanical wave equations on metric graphs, for which the
boundary conditions at the branching points (vertices) and bond ends are
imposed. The metric graph itself is determined as a set of bonds with assigned
length and which are connected to each other at the vertices according to a
rule called the topology of a graph, which is given in terms of the adjacency
matrix~\cite{Uzy1,Uzy2}:
\begin{equation*}
C_{ij}=C_{ji}=
      \begin{cases}1 & \text{ if }\; i\;
     \text{ and }\; j\; \text{ are connected, }\\
      0 & \text{ otherwise, }\end{cases}
\end{equation*}
for $i,j=1,2,\dots,N$.

Here, following Ref.\ci{Bolte}, we briefly give the general description of the
Dirac equation on quantum graphs. For the sake of simplicity we will do this
for a star graph. However, extension to an arbitrary graph is rather trivial.
Let's consider the Dirac equation (in the units $\hbar=m=c=1$) on the star
graph with $N$ bonds with finite lengths, $b_j\sim[0,L_j],\,j=1,2,...,N$, given
by
\begin{equation}\label{de00}
     \mathcal{D}\psi_j= E\psi_j,
\end{equation}
where $\psi_j=(\phi_j, \chi_j)^\top$, and the Dirac operator is
given as
\begin{equation}\label{eq16}
    \mathcal{D}:=-i\sigma_y\partial_x+\sigma_z,
\end{equation}
with $\sigma_y$ and $\sigma_z$ being the Pauli matrices:
$$\sigma_y=\begin{pmatrix}0&-i\\i&0\end{pmatrix}\,\,\,\,\,\text{ and }\,\,\,\,\,
\sigma_z=\begin{pmatrix}1&0\\0&-1\end{pmatrix}.$$

To solve Eq.\re{de00}, one needs to impose the boundary conditions at the
vertices. Such boundary conditions should keep the Dirac operator on a graph as
self-adjoint. In the case of the Schr\"odinger equation on graphs general
boundary conditions providing the self-adjointness have been derived in
\ci{Kost}. We introduce the following skew-Hermitian bilinear quadratic form
for the above star graph:
\begin{equation}\label{skewh1}
\begin{split}
\Omega(\psi,\varphi)=&\langle \mathcal{D}\psi, \varphi\rangle-\langle \psi, \mathcal{D}\varphi\rangle\\
=&\underset{j=1}{\overset{N}{\sum}}\left[\phi_j(0)v_j^*(0)-\phi_j(L_j)v_j^*(L_j)\right.\\
&-\left.\chi_j(0)u_j^*(0)+\chi_j(L_j)u_j^*(L_j)\right],
\end{split}
\end{equation}
where $\psi=\left(\psi_1,\psi_2,...,\psi_N\right)$,
$\varphi=\left(\varphi_1,\varphi_2,...,\varphi_N\right)$, and
$\psi_j=\left(\phi_j,\chi_j\right)^\top$,
$\varphi_j=\left(u_j,v_j\right)^\top$.

Then one can prove (see, \ci{Bolte}) that the self-adjointness of the Dirac
operator on graph is provided by the following requirement:
\begin{equation}\label{skew01}
\Omega(\psi,\varphi)=0.
\end{equation}

A set of vertex boundary conditions fulfilling this requirement can be written
as
\begin{equation}\label{bc1}
\begin{split}
&\phi_1(0)=\phi_2(0)=...=\phi_N(0),\\
&\chi_1(0)+\chi_2(0)+...+\chi_N(0)=0,\\
&\phi_1(L_1)=\phi_2(L_2)=...=\phi_N(L_N)=0.
\end{split}
\end{equation}
The choice of the vertex coupling comes from the physical motivation, i.e., from their relevance to real physical systems appearing in condensed matter physics.

The general secular equation for finding the eigenvalues, $k_n$, which are
derived from the boundary conditions \re{bc1}, can be written as \ci{Harrison}
\begin{equation}\label{seceq1}
\det\left(\begin{array}{cc}{\bf A}&{\bf
B}\\e^{ik\mathbf{L}}&e^{-ik\mathbf{L}}\end{array}\right)=0,
\end{equation}
where
$${\bf A}=\left(\begin{array}{cc}{\bf 1}_{N-1}^T&-{\bf I}_{N-1}\\1&{\bf
1}_{N-1}\end{array}\right),
$$
$${\bf B}=\left(\begin{array}{cc}{\bf 1}_{N-1}^T&-{\bf I}_{N-1}\\-1&-{\bf
1}_{N-1}\end{array}\right),
$$
with ${\bf 1}_{N-1}=\underbrace{(1,1,...,1)}_{N-1}$, ${\bf I}_{N-1}$ is
$(N-1)\times(N-1)$ the identity matrix, and
$e^{ik\mathbf{L}}=\mathrm{diag}\left(e^{ikL_1},e^{ikL_2},...,e^{ikL_N}\right)$.

For positive energies $E$ the general solution of Eq.\re{de00} can be written
in the form of plane waves as
\begin{equation}\label{sol1}
\psi_j(x,k)=\mu_j(k)\left(\begin{array}{cc}1\\i\gamma(k)\end{array}\right)e^{ikx}+\hat{\mu}_j(k)\left(\begin{array}{cc}1\\-i\gamma(k)\end{array}\right)e^{-ikx}
\end{equation}
with $k>0$ and
\begin{equation}\label{koef1}
\gamma(k):=\frac{E-1}{k},\,E=\sqrt{k^2+1}.
\end{equation}
The coefficients $\mu_j(k)$ and $\hat{\mu}_j(k)$ are determined by imposed
boundary conditions that provide self-adjointness of $\mathcal{D}$.

Similarly to that in \ci{Uzy1}, one can define also the bond scattering matrix,
$\mathbf{S} =S_{(ij)(lm)}$, which describes scattering of the Dirac particle on
bond $(ij)$ to that on bond $(lm)$ ($(ij)$ and $(lm)$ must be connect at $m$
\ci{Bolte}). The secular equation in terms of the scattering matrix can be
written as \ci{Bolte}
$$
\det(\mathbf{I}-\mathbf{S})=0.
$$
Spectral (with the focus on quantum chaos) and scattering properties of Dirac
particles on graphs have been studied in \ci{Bolte} using the above secular
equations and properties of the $S$-matrix. Also, the analysis of the
time-reversal symmetry in the transmission matrix and the properties of the
trace formula have been considered in \ci{Bolte}. A zeta function based study
of the Dirac operator on graphs was presented in \ci{Harrison}.

\begin{figure}[t!]
\includegraphics[width=9cm]{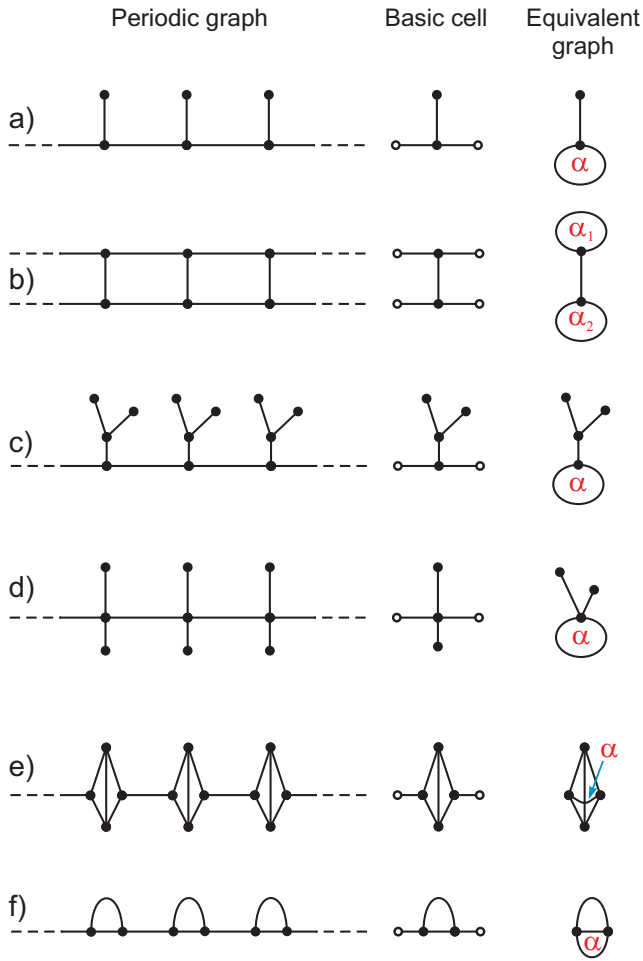}
\caption{Some types of periodic graphs with their basic cells and corresponding equivalent  graphs.}\label{fig2}
\end{figure}

\begin{figure}[t!]
\includegraphics[width=7cm]{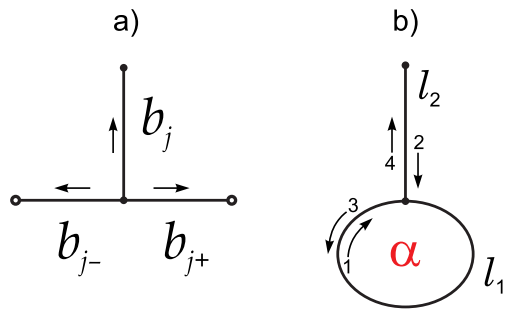}
\caption{(a) Mapping of the $j$th basic cell of the comb graph on to its
equivalent, (b) a loop vested with a magnetic flux . Arrows show directions of
the assigned coordinates.}\label{fig3}
\end{figure}

\section{Dirac particles on periodic quantum graphs}

Here, using an extension of the approach, developed in \ci{Grisha1}, we
consider a Dirac equation on a periodic graph by computing band spectra of
different topologies. Within such an approach, one can treat a wide variety of
periodic graphs. Although one can construct different types of periodic graphs
using simple graphs as ``unit cells,'' we will focus on the types presented in
Fig.~\ref{fig2}. In our approach, a periodic graph is considered as a repeating
structure of basic graphs, which can be called unit cells, i.e. the ``graph
lattice'' is a periodic or quasiperiodic structure of basic graphs.  We assume
that in all cases, a periodic or quasi-periodic graph can be ``mapped'' onto
the simple graphs, which (for each graph) are shown in the last column in
Fig.~\ref{fig2}. We note that there is well developed  and powerful approach
for the treatment of spatially periodic quantum systems, which is known under
different names, such as the Bloch method, the Floquet method, and the Gelfand
transformation. Within such an approach, the prescription for solving the Dirac
equation on periodic graph can be formulated as follows:
\begin{enumerate}[label=\roman*)]
\item Write Dirac equation on each bond of the basic cell;
\item Impose the vertex boundary conditions for each node of the basic cell;
\item Impose the ``intercell'' boundary conditions for the whole periodic graph;
\item Derive the secular equation in terms of the scattering matrix from the vertex and intercell boundary conditions;
\item Construct the scattering matrix and find eigenvalues from the secular equation.
\end{enumerate}
Below we demonstrate the application of this prescription for the periodic comb
graph presented in Fig.~\ref{fig2}(a). We define the bonds of the graph as
$b_{j\pm}\sim(0,l_1/2),\,b_{j}\sim(0,l_2),\,j=...,-1,0,1,...$ (see,
Fig.~\ref{fig3}). Directions of assigned coordinates on each bond are shown by
arrows in Fig.~\ref{fig3}(a). To each bond $b_j$ of the graph we assign a
coordinate $x_j$, which indicates the position along the bond: for bond $b_1$
it is $x_j\in [0,l_1/2]$. One can use the shorthand notation $\Psi_j(x)$ for
$\Psi_j(x_j)$ and it is understood that $x$ is the coordinate on the bond $j$
to which the component $\Psi_j$ refers.

On each bond of this graph we have the following one-dimensional Dirac equation
(in the units $\hbar=m=c=1$):
\begin{equation}\label{de1}
\begin{split}
-\partial_x\chi_b + \phi_b&=E\phi_b,\\
\partial_x\phi_b - \chi_b&=E\chi_b,
\end{split}
\end{equation}
for $b\in\{b_{j-},b_{j+},b_j\}$.

The vertex boundary conditions for each basic cell are imposed as
\begin{align}
\phi_{b_{j-}}(0)&=\phi_{b_{j+}}(0)=\phi_{b_j}(0),\label{vbc1}\\
\chi_{b_{j-}}(0)&+\chi_{b_{j+}}(0)+\chi_{b_j}(0)=0,\label{vbc2}\\
\chi_{b_j}(l_2)&=0.\label{vbc3}
\end{align}
For the periodic graph presented in Fig.~\ref{fig2}(a), one can impose the
following quasiperiodic (inter-cell) conditions:
\begin{align}
\phi_{b_{j+}}(l_1/2)&=e^{i\alpha}\phi_{b_{j-}}(l_1/2),\label{qpbc1}\\
\chi_{b_{j+}}(l_1/2)&=-e^{i\alpha}\chi_{b_{j-}}(l_1/2).\label{qpbc2}
\end{align}
Vertices with these boundary conditions are denoted by the empty
circles in Figs.~\ref{fig2} and~\ref{fig3}. It can be shown that the
boundary conditions \re{qpbc1} and \re{qpbc2} do not break the
self-adjointness of the Dirac operator on graph, since they are
consistent with Eq.\re{skew01}

A general solution of the system of Eq. (\ref{de1}) can be written as
\begin{equation}\label{sol}
\left(\begin{array}{cc}\phi_b(x)\\
\chi_b(x)\end{array}\right)=\mu_b\left(\begin{array}{cc}1\\i\gamma\end{array}\right)e^{ikx}+\hat{\mu}_b\left(\begin{array}{cc}1\\-
i\gamma\end{array}\right)e^{-ikx}.
\end{equation}
For the whole periodic graph in Fig.~\ref{fig2}(a), for the direction 1 [see
Fig.~\ref{fig3}(b)] the general solution can be written in terms of the
following outgoing and incoming waves at the vertex:
\begin{equation}\label{newf1}
\begin{split}
\left(\begin{array}{cc}\phi_1(l_1-x)\\
\chi_1(l_1-x)\end{array}\right)&=e^{i\alpha}\left(\begin{array}{cc}\phi_{b_{j-}}(x)\\
-\chi_{b_{j-}}(x)\end{array}\right),\\
\left(\begin{array}{cc}\phi_1(x)\\
\chi_1(x)\end{array}\right)&=\left(\begin{array}{cc}\phi_{b_{j+}}(x)\\
\chi_{b_{j+}}(x)\end{array}\right).
\end{split}
\end{equation}
From the periodic boundary conditions given by Eqs.(\ref{qpbc1})-(\ref{qpbc2}) and Eq.(\ref{sol}) we have
\begin{align}
\left(\begin{array}{cc}\phi_1(y)\\
\chi_1(y)\end{array}\right)&=\mu_1^{(1)}e^{i\alpha+ikl_1}\left(\begin{array}{cc} 1\\
-i\gamma\end{array}\right)e^{-iky}\nonumber\\
&+\hat{\mu}_1^{(1)}e^{i\alpha-ikl_1}\left(\begin{array}{cc}1\\
i\gamma\end{array}\right)e^{iky},\frac{l_1}{2}\leq y \leq l_1,\label{sol1}\\
\left(\begin{array}{cc}\phi_1(y)\\
\chi_1(y)\end{array}\right)&=\mu_1^{(2)}\left(\begin{array}{cc}1\\
i\gamma\end{array}\right)e^{iky}+\hat{\mu}_1^{(2)}\left(\begin{array}{cc}1\\
-i\gamma\end{array}\right)e^{-iky},\nonumber\\
&0\leq y \leq \frac{l_1}{2},\label{sol3}
\end{align}
where
\begin{equation}\label{mu13}
\hat{\mu}_1^{(1)}=\mu_1^{(2)}e^{-i\alpha+ikl_1},\,\hat{\mu}_1^{(2)}=\mu_1^{(1)}e^{i\alpha+ikl_1}.
\end{equation}
For the direction 4 we have
\begin{align}
\left(\begin{array}{cc}\phi_4(x)\\
\chi_4(x)\end{array}\right)&=\mu_4\left(\begin{array}{cc}1\\
i\gamma\end{array}\right)e^{ikx}+\hat{\mu}_4\left(\begin{array}{cc}1\\
-i\gamma\end{array}\right)e^{-ikx}.\label{sol2}
\end{align}
From the boundary conditions (\ref{vbc1})-(\ref{vbc3}) we have
\begin{equation}\label{mus}
\begin{split}
\mu_1^{(1)}&+\hat{\mu}_1^{(1)}=\mu_1^{(2)}+\hat{\mu}_1^{(2)}=\mu_4+\hat{\mu}_4,\\
\mu_1^{(1)}&-\hat{\mu}_1^{(1)}+\mu_1^{(2)}-\hat{\mu}_1^{(2)}+\mu_4-\hat{\mu}_4=0,\\
\mu_4&e^{ikl_2}=\hat{\mu}_4e^{-ikl_2}.
\end{split}
\end{equation}

From Eqs.(\ref{mu13}) and (\ref{mus}) we obtain the following
secular equation for finding the eigenvalues of the relativistic spin-half quasiparticle on periodic quantum graph:
\begin{equation}\label{seceq1}
F(k;\alpha):=\det(\mathbf{I}-\mathbf{S})=0,
\end{equation}
where $\mathbf{I}$ is the four dimensional identity matrix and $\mathbf{S}$ is
the scattering matrix, which depends on the graph topology. For the comb graph
in Fig.~\ref{fig2}(a), the explicit form of $\mathbf{S}$ can be written as
$$
\mathbf{S}=\left(\begin{array}{cccc}
e^{-i\alpha+ikl_1}&0&e^{-i\alpha}-e^{ikl_1}&0\\
-1+e^{-i\alpha+ikl_1}&0&e^{ikl_1}-e^{-i\alpha}&e^{ikl_2}\\
-e^{ikl_1}&e^{i\alpha}&0&e^{i\alpha+ikl_2}\\
0&e^{ikl_2}&0&0
\end{array}\right).
$$

\begin{figure}[t!]
\includegraphics[width=8.7cm]{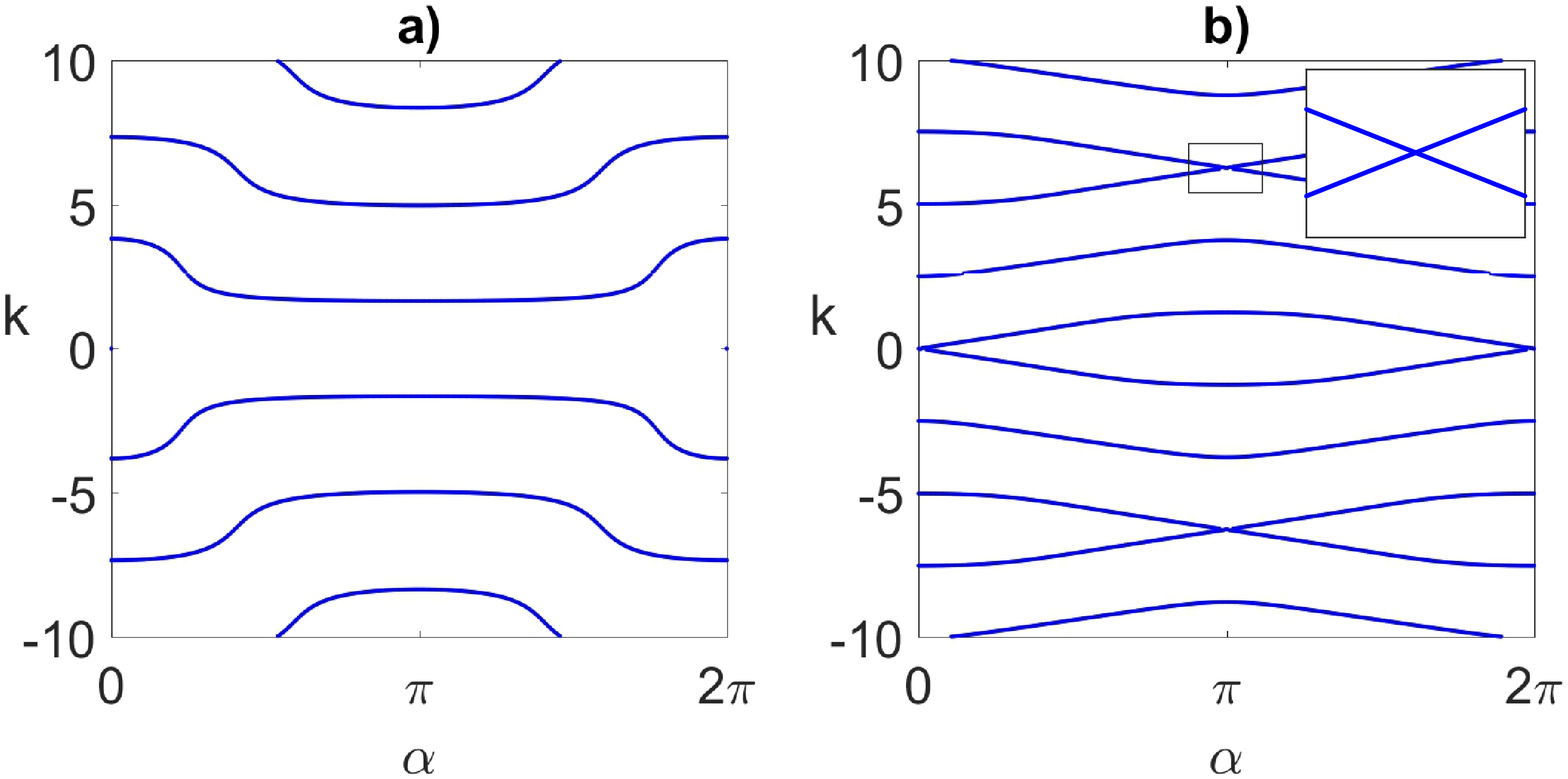}
\caption{Band spectra of a Dirac particle on the periodic  comb graph
(presented in Fig.~\ref{fig3}). Lengths are fixed by the relation $l_1=\beta
l_2$, where $l_2$ is unit length ($l_2=1$) for the values $\beta=0.2$ (a) and
$\beta=2.5$ (b).}\label{fig4}
\end{figure}

\section{Band spectra of the Dirac particles  on periodic quantum graphs}

An important characteristic of the periodic quantum structure is its band
spectrum, which is the eigenenergy spectrum as a function of a system
parameter. It characterizes most of the electronic properties, including
electric conductance. Usually, for spatially periodic systems the band
structure is symmetric with respect to the gap lying between the conductance
and valence bands. This concerns also periodic branched structures described in
terms of periodic quantum graphs. The periodicity of the graph is considered
here with respect to basic cells. From the practical viewpoint, it is important
to study dependence of the band spectra on the topology of a periodic graph and
from different metric parameters of the basic cell. The latter implies the
question how does the band spectrum change by changing the length of bonds
connecting basic cells ($l_1$ in Fig.~\ref{fig3})?

\begin{figure}[t!]
\includegraphics[width=8.7cm]{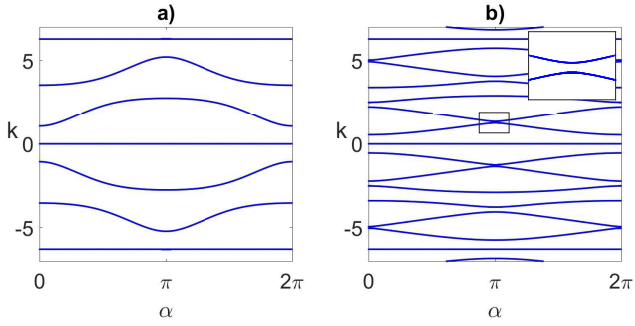}
\caption{Band spectra of a Dirac particle on the periodic  ladder graph. Lengths are fixed by the relation $l_1=\beta l_2$, where $l_2$ is unit length ($l_2=1$) for the values $\beta=0.2$ (a) and $\beta=2.5$ (b).}\label{fig5}
\end{figure}

\begin{figure}[t!]
\includegraphics[width=8.7cm]{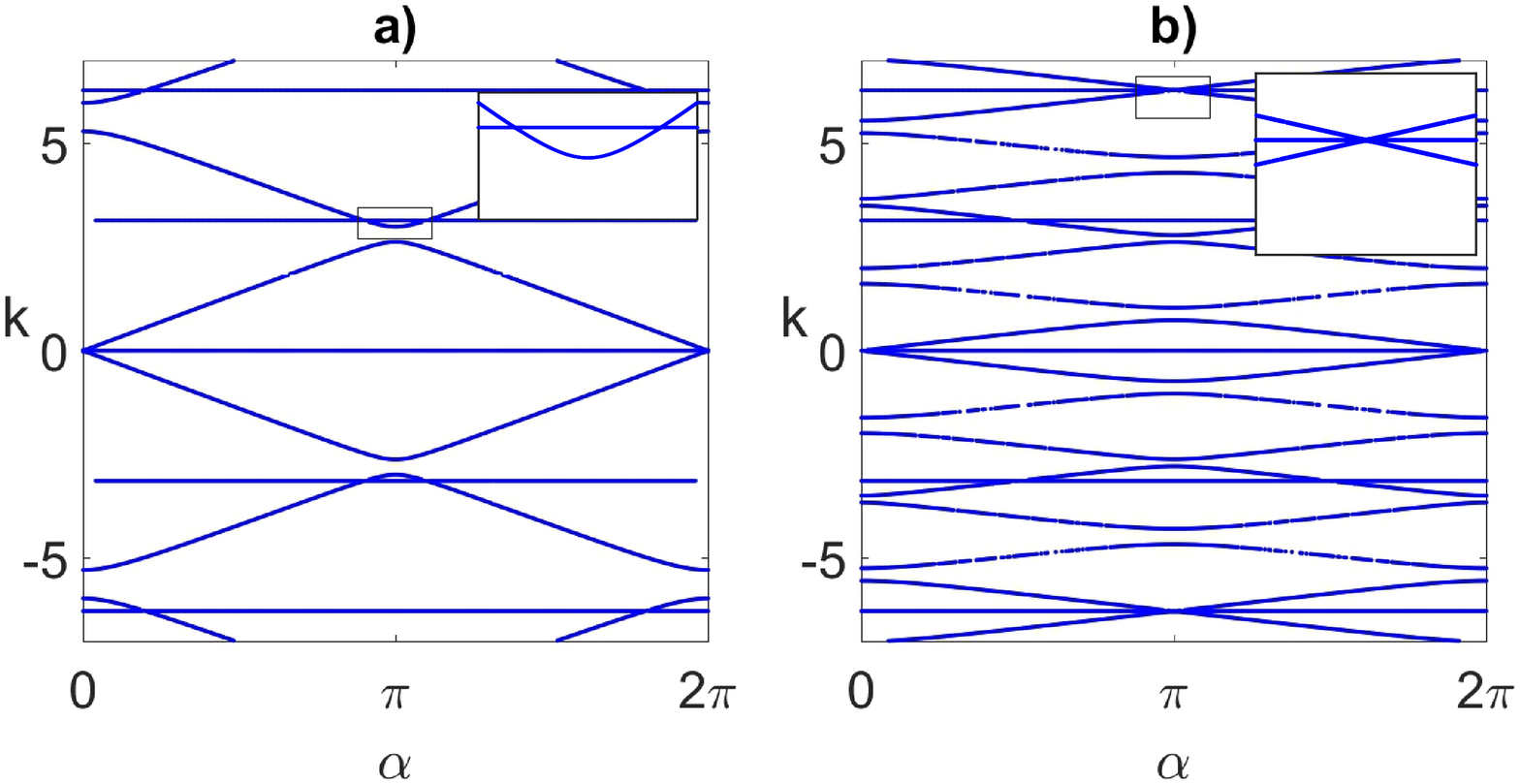}
\caption{Band spectra of a Dirac particle on the periodic loop graph. Lengths are fixed by the relation $l_1=\beta l_2$, where $l_2$ is unit length ($l_2=1$) for the values $\beta=0.2$ (a) and $\beta=2.5$ (b).}\label{fig6}
\end{figure}

Thus, using the secular equation~\re{seceq1} that holds true for an arbitrary
periodic quantum graph, we calculate the band spectrum of a Dirac particle on
the comb graph. We study the behavior of the band spectrum by fixing the length
of $l_2$ as a unit and changing the length of $l_1$ given via the relation
$l_1=\beta l_2$.

Figure~\ref{fig4} presents the band spectra of the comb graph [see
Fig.~\ref{fig2}(a)] plotted for two values of $\beta$: $\beta =0.2$ and $\beta
=2.5$. One can observe that the gap between the levels decreases when $\beta$
increases, which means increasing the distance between ridges. Moreover, one
can see level crossing (see, insets).

Similarly, one can calculate the band spectra for periodic ladder and loop
graphs, presented in Figs.~\ref{fig2}(b) and~\ref{fig2}(f), respectively. For
the sake of simplicity, for the ladder graph we choose
$\alpha_1=\alpha_2=\alpha$, and for the loop graph we choose the equal lengths
of the loop and the bond connecting the loop ends. The scattering matrices for
these graphs can be written as
\begin{widetext}
$$
\mathbf{S}_\text{ladder}=\left(\begin{array}{cccccc}
e^{-i\alpha+ikl_1}&0&0&e^{-i\alpha}-e^{ikl_1}&0&0\\
0&e^{-i\alpha+ikl_1}&0&0&e^{-i\alpha}-e^{ikl_1}&0\\
1&0&0&e^{ikl_1}&0&-e^{ikl_2}\\
-e^{i\alpha}+e^{ikl_1}&0&e^{i\alpha}&e^{i\alpha+ikl_1}&0&e^{i\alpha+ikl_2}\\
0&-e^{i\alpha}+e^{ikl_1}&e^{i\alpha+ikl_2}&0&e^{i\alpha+ikl_1}&-e^{i\alpha}\\
0&1&-e^{ikl_2}&0&e^{ikl_1}&0
\end{array}\right),
$$
$$
\mathbf{S}_\text{loop}=\left(\begin{array}{cccccc}
0&1&0&-e^{ikl_1}&e^{ikl_2}&0\\
0&0&e^{-i\alpha+ikl_3}&0&-e^{ikl_2}&e^{-i\alpha}\\
e^{ikl_1}&0&0&1&0&-e^{ikl_3}\\
-e^{ikl_1}&e^{ikl_2}&0&0&1&0\\
e^{-ikl_2}&e^{-ikl_2}&-e^{-i\alpha+ikl_3-ikl_2}&-e^{ikl_1-ikl_2}&0&e^{-i\alpha-ikl_2}\\
-e^{ikl_1-ikl_3}&-e^{ikl_2-ikl_3}&e^{-ikl_3}&e^{-ikl_3}&e^{-ikl_3}&0
\end{array}\right),
$$
\end{widetext}
where $l_2=l_3$ according to our assumption above.

Figure~\ref{fig5} presents band spectra for the periodic ladder graph for the
same values of $\beta$ and $l_2$ (the length of ladder steps). The spectral
picture is the same as that for the comb graph, the gap between the levels
decreases when $\beta$ increases. But the band spectrum manifests avoided
crossing of the levels (see inset).

In Fig.~\ref{fig6} plots of the band spectra for periodic loop graph are
presented for the same values of parameters as those for the comb and ladder
graphs. For both values of $\beta$ a crossing of levels can be observed.

This study shows that the electronic properties of the periodic quantum
structure strongly depend on the system parameter $\beta$, which in our case is
the distance between basic cells.

\begin{figure}[t!]
\includegraphics[width=8cm]{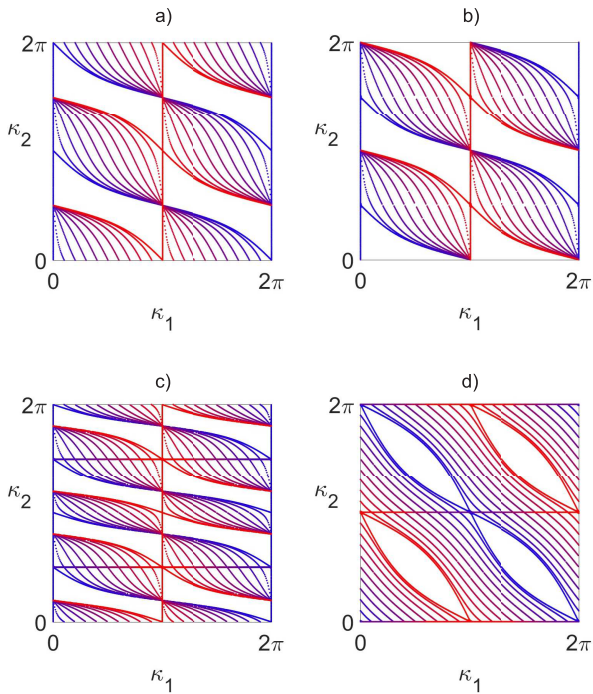}
\caption{(color online) The zero sets of $\Phi(k_1,k_2;\alpha)$ for a range of
values of $\alpha\in[0,\pi]$ for (a) the periodic comb, (b) ladder, (c)
periodic graph shown in Fig.~2(c) and (d) loop graphs  using color scale from
blue ($\alpha=0$) to red ($\alpha=\pi$).}\label{fig7}
\end{figure}

\section{Probability to be in the spectrum}

A remarkable result of the Ref. \ci{Grisha1}, where a nonrelativistic
counterpart of our problem is considered, is revealing the universal behavior
of the probability that a randomly chosen momentum belongs to the spectrum of
the periodic graph. Namely, the probability to be in the spectrum does not
depend on the edge lengths and is also invariant within some classes of graph
topologies. The basic cell classes, which manifest such a behavior are the
decorations that attach to the base line by means of a single edge, e.g. as in
Figs.~2(a)-2(c). Such behavior cannot be observed, for instance, in cases of
decorations presented in Figs.~2(d)-2(f). Investigating such a probability and
existence of universality properties for a relativistic case should be
interesting both from fundamental as well as practical viewpoints.

The same trick used in \ci{Grisha1} (but originally belonging to Barra and
Gaspard \ci{Barra}) can be directly applied to our case. To calculate the band
spectrum of the considered periodic quantum graph: we introduce a new function
\begin{equation}\label{secfun}
\Phi(\kappa_1=kl_1,\kappa_2=kl_2;\alpha):=F(k;\alpha),
\end{equation}
where $\kappa_1$ and $\kappa_2$ need only be known modulo $2\pi$. In this way, for a fixed $\alpha$ we found solutions of
\be
\Phi(\vec{\kappa};\alpha)=0,
\lab{sfunct}
\ee
where $\vec{\kappa}(k)=k\cdot(l_1,l_2)\mod 2\pi$ and $k$ belongs to the spectrum of the periodic graph.

Figure~\ref{fig7} presents dependence of $k_2$ on $k_1$ at different values of
$\alpha$ calculated using Eq.~\re{sfunct} for the four periodic graph
topologies depicted in Figs.~2(a)-2(c) and 2f.

Using the same  approach used in \ci{Grisha1}, we calculate the probability
$p_\sigma$ for a random $k$ to be in the spectrum $\sigma$. However, unlike the
nonrelativistic counterpart where probability can be estimated analytically
\ci{Grisha1}, in the relativistic case the secular equation has very
complicated form. Therefore one should compute the probability numerically.
Using the symmetry properties of the spectrum plotted in Fig.~\ref{fig7} one
can calculate the probability by finding the area of $1/8$-th of the zero sets,
which is the part in the lower left corner, bounded by the coordinate axes and
the function $\kappa_2=\varphi(\kappa_1;\pi)$. The numerical calculations of
the probability to be in the spectrum show that for (a) comb and (b) ladder
graphs and (c) the graph shown in Fig.~2(c) it takes the (same) value
$p_\sigma\approx0.64$, while for (d) the loop graph the probability is
different, $p_\sigma\approx0.73$. A similar situation was observed for the
nonrelativistic counterpart considered in  \ci{Grisha1}. Thus, for the
relativistic case described by the Dirac equation on periodic graphs we
observed similar universality for the decorations that attach to the base line
by means of a single edge (at least for the topologies presented in Fig.~2).

\section{Conclusions}
We studied the Dirac equation on periodic quantum graphs with a focus on the
eigenvalue problem. The secular equation allowing us to find a band spectrum of
Dirac quasiparticles in networks is derived from quasiperiodic boundary
conditions. It is shown that these latter do not break the self-adjointness of
the Dirac operator on graphs. Band spectra of different periodic graphs are
computed. Universality of the probability to be in the spectrum for certain
graph topologies is shown by numerical calculations. The above model can find
applications in the study of electronic properties of different
quasi-one-dimensional branched (periodic) structures, such as, e.g.,
periodically branched conducting polymers, where transport of polarons can be
described in terms of the Dirac equation on graphs.

\end{document}